# Agent-based Model Construction In Financial Economic System

**Hokky Situngkir[1]**
(hokky@elka.ee.itb.ac.id)
Dept. Computational Sociology
Bandung Fe Institute

**Yohanes Surya[2]**
(yohaness@centrin.net.id)
Dept. Physics
Universitas Pelita Harapan

[1] *Board of Science* Bandung Fe Institute, http://www.geocities.com/quicchote
[2] Professors in Dept. Physics Universitas Pelita Harapan dan *Board of Advisory* Bandung Fe Institute,
http://www.ekonofisika.com

## Abstract

The paper gives picture of enrichment to economic and financial system analysis using agent-based models as a form of advanced study for financial economic data post-statistical-data analysis and micro-simulation analysis. Theoretical exploration is carried out by using comparisons of some usual financial economy system models frequently and popularly used in econophysics and computational finance. Primitive model, which consists of agent micro-simulation with fundamentalist strategy, chartist, and noise, was established with an expectation of adjusting micro-simulation analysis upon stock market in Indonesia. The result of simulation showing how financial economy data resulted analysis using statistical tools such as data distribution and central limit theorem, and several other macro-financial analysis tools previously shown (Situngkir & Surya, 2003b). This paper is ended with several further possible advancements from the model built.

## 1. Background

Within the previous paper (Situngkir & Surya, 2003a), it has been shown how multi-agent analysis has given new atmosphere to analyze financial economy system which provides thousands to millions transaction data. This gives inspirations and ways for the genesis of econophysics that is rich with quantitative analysis. There has been a revolutionary paradigm shift within financial analysis with the involvement of many physicists in the works of financial analysis. Furthermore, the developing computational technology that forges analysis from any disciplines to use multi-agent analysis as one of its analytical tools for giving better understanding of social system, in this case, financial system.

Complexity concept, in which inside lies various propositions of emergence, macro-micro linkage factors, has brought multi-agent analysis to be one of computational devices that is accountable for explaining various events and measurable events at macro level within micro structure that brings them to social sciences (Macy & Willer, 2002). In the advancement on financial system analysis developed with various statistical mechanic models (econophysics) is however giving out many enigmatic facts. A perspective of multi-agent analysis is hopefully able to explain it. As shown in Figure 1, time-series data analysis giving a complete description of financial economy data properties – yet, in the other side various argument of the micro properties of financial economy agents tried to seek a linkage between macro analysis (time-series data) e.g.: volatility grouping, Levy distribution, multifractality, along with the behavior of economy agents. This is a bridge given by multi-agent analysis: finding connection between micro-simulation properties of financial system with its macro or statistical properties.

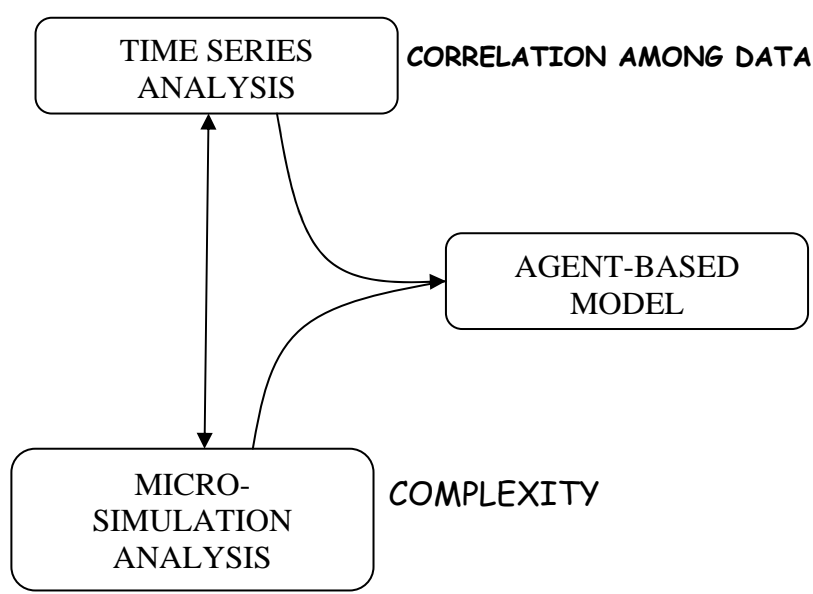

**Figure 1**

The position of agent-based modeling as an analytical bridge of statistical properties of financial data with micro-simulation properties that causing it.

How to explain things in macro-level in the situation of micro-simulation is the main motivation of using multi-agent simulation analysis in social sciences, including financial economy (Axtell, 2000). Several common examples on the use of multi-agent models in financial economy analysis (be it stock market or foreign exchange market) are Bak, et.al. (1996), Takahashi & Terano (2003), Zimmeman, Neuneier, & Grothman (2001), King, Streitchenko, & Yesha, (20030, while there is also endeavor to pronouncing macro-economy in multi-agent language (Bruun, 1999), and investor's behavior (Farmer, 2001).

This paper is trying to give illustrations of multi-agent models that have been frequently utilized in finance, continued with building primitive model from financial economy system with adaptive efforts upon stock market in Indonesia and the direction and possible development to farther analysis level.

## 2. Agent-based Modeling in Stock Market Analysis

On January 9[th], 2004, Jakarta Composite Index (IHSG) reached the highest point for the last five years and several analyses easily appeared to excuse the fact as the enthusiasm of stock market in the opening year. In fact, every body knows that it is too difficult to assure exactly how the thing could happen in different description levels in economy. But apparently, this frequently occurs in financial system. And so does at the bombing action in Mariott Hotel where index was immediately corrected, and several other macro-financial quantitative effects that is too difficult to explain in the level of individual economy player.

In econophysics, the fulfilled "power-law distribution" at almost all financial economy data, volatility grouping, etc has become a scientific quest that is urgently needed to answer. Multi-agent analysis is one of alternative solution emerging to solve those various puzzles since it can bridge macro and micro factor in a system.

As explained in Situngkir & Surya (2003a), there have been several multi-agent and computational models used to explain generally behavior of financial system. The most popular model is The Santa Fe Institute's Artificial Stock Market, a financial economy model that uses various micro-simulation model inspirations such as Oxford model (minority games), El Farol's bar model, etc (LeBaron, 2002). In short, this economy model tries to explain various statistical facts found in financial economy data to have understanding of the micro cause, which is in individual level of market player (Lux, 2000).

In Situngkir & Surya 2003b), we obtained three unique and interesting characters from financial economy time series data in Indonesia – that in reality have common character and suits the finding in financial economy data in several places outside Indonesia, they are:

✓ Volatility Grouping character
✓ Excess kurtosis character and fat tail distribution, or known as power-law distribution properties, and
✓ Multifractality character

To develop multi-agent model of financial system is to give explanation how can those data emerge from micro-structure of the stock market Indonesia.

In the paper, we use three general strategies from investors who become the player in financial economy Indonesia, that is fundamentalist strategy – a strategy that has tendency to maintain price in a certain index value, chartist strategy – a strategy that intent to take profit from time series data trend analysis, and at last noisy analysis – a random strategy from investor yet its influence to the whole system is relatively small.

# 3. Model Construction

The basic principle of multi-agent system is to arrange micro-rule of the system by modeling the agent's interaction pattern in order to computationally show several macro characters from the system in aggregate. In our preliminary model, the Inter-agent interaction will be monitored by a market maker, in reality, say it stock exchange who becomes the note-taker at the meeting of seller (supply) and buyer (demand) in the stock transaction. Below, explained stock market we use that is inspired and adopted from Castiglione (2001:88-104) and Marchesi, *et al.* (2000) models.

### 3.1 Market Model

In our model, we regard market as the provider of the stock buyers and sellers meeting whose own-*quote* to respond market situation at the previous time. We use market space, which is defined time (of iteration) vectors that in each time-step the market maker notes the quoted prices and then decide to whom the seller and buyer will transact – The price at the done transaction is then said to be the market-price. These data price will then be analyzed with statistical analysis. For the sake of simplicity, we involve only one stock in our market model, and the price ( $p_t$ ) is stipulated by the single "market maker". So, after agent gives choice of sell, buy, or stay in-active, the market maker gives value $p_{t+1}$ as excess demand function followed:

$$D_t = \sum_i x_t^{(i)} \qquad \ldots\ldots\ldots\ldots(1)$$

while the price difference proportional to the excess demand, as:

$$p_{t+1} - p_t \sim D_t \qquad \ldots\ldots\ldots\ldots(2)$$

Responding every order, selling or buying, the market will only serves transaction in certain minimum slot of the.

### 3.2   Economic Agent

There are three agent attributes that always change along simulation and modeling operation, they are:

✓ Choices of sell, in-active, or sell, represented as $x_t^{(i)} \in \{-1,0,1\}$

✓ Stock or capital that will be invested in stock market represented as $c_t^{(i)}$

✓ Number of stocks that become investment in stock market $n_t^{(i)}$. So that, in each iteration, the total asset of each agent: $k_t^{(i)} = n_t^{(i)} p_t + c_t^{(i)}$

Meanwhile, each agent also acquires fixed attributes that become the basic and background as investor, i.e.:

■ Strategy. Each agent has certain strategy determining its stand $x_t^{(i)} \in \{-1,0,1\}$. For simplification, we will monitor three basic strategies, they are:

– *Fundamentalist strategy*, a strategy that always have tendency to hold a price at a certain value. Means it will sell $x_t^{(i)} = -1$ if the price is higher than its fundamental value and vice versa buy for price lower than its fundamental value.

– *Chartist strategy*, known as strategy for those who monitor market trend for certain history referred horizon ($x_t^{(i)} = -1$) – this method also known as moving average (MA). Agent sells if MA value:

$$\overline{m_t}(h) = \frac{1}{h} \sum_{t'=t-h}^{t-1} p_{t'} \qquad \ldots(3)$$

computed with h time horizon is larger than the price:

$$p_t^+(\delta) = p_t + p_t \delta \qquad \ldots(4)$$

where $\delta = (0,1)$ as input parameter. As shown in Figure 4, they will sell if the value of MA parameter is below the price: $p_t^-(\delta) = p_t - p_t \delta$

– *Noisy strategy*. Choosing transaction actions of selling randomly with probability 0,5 but only buy if he feels save to sell. Average value of price they paid for the whole stock they own in macro relatively small as $p_t$:

$$\frac{1}{n_t^{(i)}} \sum_{j=1}^{n_t^{(i)}} p_{t_j} < p_t \qquad \ldots (5)$$

where $t > 1$ and $t_j < t, \forall j$

■ Influence strength. If we run simulation with the above parameters *par excellence* then we can find that the price fluctuation will behave Gaussian distribution – means that it will be unrealistic to describe statistical properties of stock price that in reality power-law distributed. Cont & Bouchaud (2000) showed that volatility clustering phenomenon is consequently the result of information diffusion or collective information among economic agents in micro-view of the system and in return appeared to be fat tail in its distribution in macro-view. Each agent affects and is affected by her surroundings on a variable of influence strength, say $s^{(i)}$. How the influence strength affect the decision to be chosen by agents? Each decision, $x_t^{(i)} \in \{-1,0,1\}$, is determined by agent's strategy - we normalize the value in the interval between $-1,0,1$. Therefore it can be seen as probability form. Then, the probability can be stated as distribution:

$$P[x_t^{(i)} = x] = \frac{\sum\limits_{j}^{x_t^{(i)}=x} s^{(j)}}{\sum\limits_{j} s^{(j)}} \qquad \ldots(6)$$

with total possibility follows:

$$\sum_x P[x_t^{(i)} = x] = 1 \qquad \ldots(7)$$

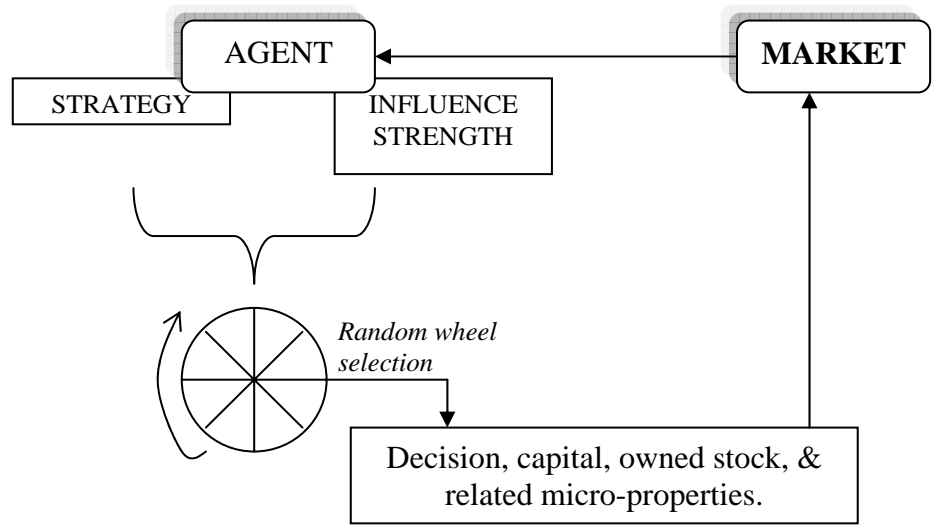

**Figure 2**
The Simulation process.

Eventually, decision of each agent, be it sell, stay in active, or buy obtained by using principles of random wheel selection. The bigger the strength of an agent's influences is, the bigger the probability to influence others to decide as she does.

**Table 1**
**Initial Simulation Configuration**

| Parameters | Value |
|---|---|
| Number of agent (investor) | 250 |
| Formation *fundamentalis-chartist-noisy* | 77-84-89 |
| Total Marketed Stock | 10.000 |
| Basic price each stock | Rp 3200,- |
| Maximum initial capital | Rp 1.000.000,- |
| Number of stock each slot | 5 |
| Distribution of initial capital, owned stock, strategy, and influence strength in uniformly random in the beginning of the simulation. | |

## 4. Simulation Results & Analysis

We involve market arrangement as shown in Table 1. We perform some experiments with certain variables varied prior to the need of analysis. The process of simulation in short is illustrated in **Figure 2**. An example resulted from simulation we ran is shown in Figure 3. In the figure, it is shown the fluctuating stock price and return value with some occurrences of herding behavior. This is the first financial economy data character that is fulfilled by our primitive multi-agent simulation result.

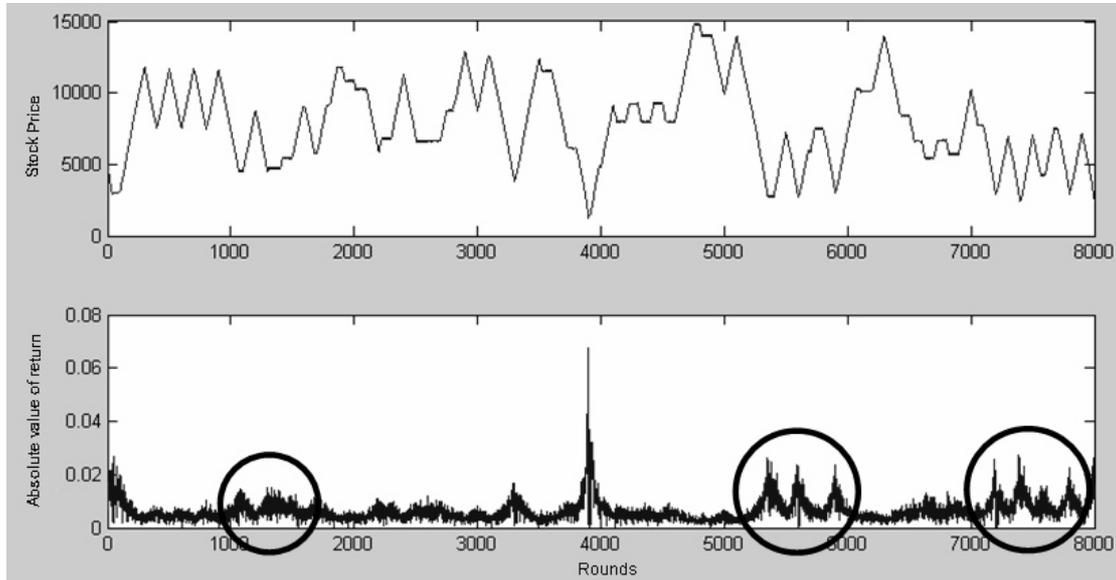

**Figure 3**
Experimental result with multi-agent simulation with basic configuration as shown in Table 1. Also shown volatility clustering on the data resulted.

The most interesting aspect from multi-agent simulation result in this financial economy system is that the resulting outcome of artificial data that follows the power-law distribution, whose strong leptokurtosis character with fat tail. This is shown in Figure 4 where return per iteration described each for negative and positive tail and both fitted with square root distribution.

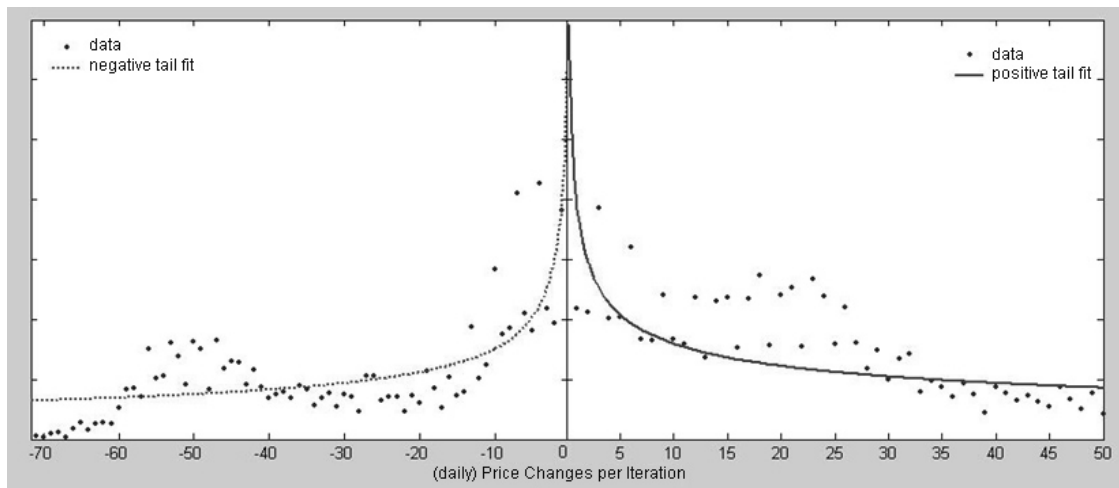

**Figure 4**
Distribution of Return showing a strong leptokurtosis character

In the experiment presented here, there is no endeavor yet to discover the formation of micro-properties (artificial economy) with the true statistical properties we gain in practices of stock exchange data in Indonesia. However, prior to the aim of the paper is to develop preliminary model that in general point resulting financial time series data with similar nature with the real data.

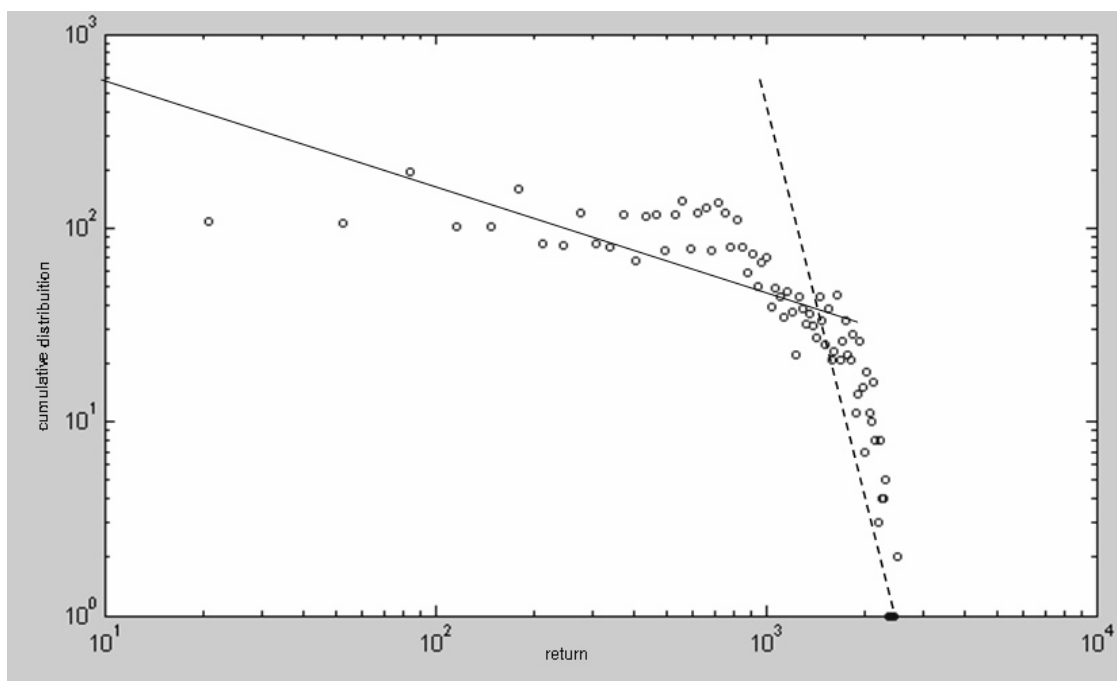

**Figure 5**
Scaling behavior of the cumulative distribution resulted

Further, figure 5 is shown scaling behavior that appeared in the existing time series data – as investigated by Stanley & Mantegna (2999: 64-75). The figure shows the slope change of data presenting the occurrence of distribution crossover from Levy to Gaussian distribution – consequently to the Central Limit Theorem.

The last property we are willing to observe is the multifractal behavior. Based on the result we obtain on scaling behavior, basically we gain general outline that implicitly express the existence of self-similarity in time series data distribution resulted by our artificial economy system. This is illustrated in Figure 6. In that picture we can obviously see the self-similarity behavior occurs from the same form of graph plot for different return value and time scaling (per 1-iteration, per-5-iteration, per-10-iteration, and per-30-iteration). It assures that the artificial financial economy system we develop can obviously generate multifractality character existed.

From the three general characters we obtained, we have seen how multi-agent analysis can be used as additional and powerful tool to find explanation from various financial time series data characters that we analyze with statistical mechanics previously. In other words, multi-agent analysis gives new pattern of theoretical exploration in econophysics approach, specifically financial physics.

# 5. Some Concluding Remarks and Further Works

Financial economy time series data character such as volatility clustering, fat tail distribution and multifractality is shown in multi-agent simulation we developed. This fact implies that by using multi-agent analysis to explain how macro-phenomenon detected by statistical mechanics that generally used in econophysics does have good prospect.

Further development of this multi-agent model nationally in Indonesia will be able to give explanation to various aspects in our financial economy system. The advantage that can be drawn is the analysis enrichment for the need of decision-making and opportunity to evaluate financial processes in stock exchange and also for the need of education by stock-exchange simulation.

As applied above, it is obvious that the use of multi-agent model gives explanation of various factors such as the relation of influence power of investor factor upon volatility clustering, and how interdependent interactions among investors comprise macro-behavior (stylized statistical facts) i.e.: multifractality and power-law distribution. Practically, within implementation of methodology proposed, we can surely have better chance and powerful tools to have evaluation of many aggregated data e.g.: indexes parameter, which in principle reflecting national stock exchange economical condition – and if necessary to re-construct for the better performance.

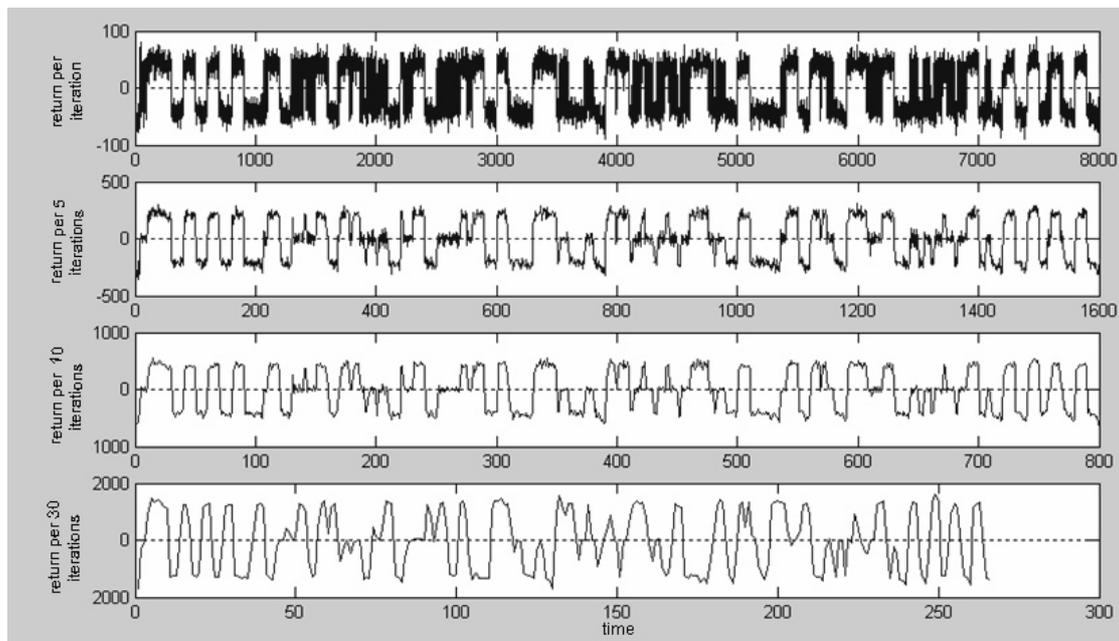

**Figure 6**

Multifractal in financial time series data resulted from multi-agent simulation

For the sake of education, this simulation can be further developed in giving information and introduction to the stock exchange system exist to common public. By developing the user-friendly programs on principles described in the paper will appropriate for this purpose.

Furthermore, several technical things from this paper need to be developed in advanced research to gain more realistic financial data are:

- Spatial techniques utilization, network, in model construction so that there will be visualization of accumulating capital from interaction of each agent.
- Adding "intelligence" to each agent so that each agent has evolutionary ability in changing techniques and strategies used in her decision-making.
- Price enumeration as the result of direct interaction between stock sellers and buyers so they can approach reality of the stock system that will be modeled and explained its various macro-quantitative factors.
- Simulation that involving several stocks or other secondary products in the market so that it can simulate stock exchange indexes. This is surely advantageous in index construction, which has reflection ability upon the better financial economy reality.

With these developments, it is hoped that we can have multi-agent model that able to be utilized in analyzing and developing the better stock system and suitable with our expectation on enthusiasm for better understanding of national economic system.

## Acknowledgements:

Research reported in this paper is carried out with financial support from the Lembaga Pengembangan Fisika Indonesia (LPFI). The writer would like to thank all partners in Bandung Fe Institute, for their critics and advices to this draft, especially Yun Hariadi for discussions on multifractality.